\documentclass[runningheads,a4paper]{llncs}
\usepackage{booktabs}
\usepackage{amssymb}
\usepackage{graphicx}
\usepackage{amsmath}
\usepackage{caption}
\usepackage{url}
\usepackage{floatrow}
\usepackage{multirow}
\newfloatcommand{capbtabbox}{table}[][\FBwidth]
\urldef{\mailsa}\path|{shuai.zhang@student., lina.yao@}unsw.edu.au|
\urldef{\mailsc}\path|{Liming.Zhu, XiWei.Xu}@data61.csiro.au|
\urldef{\mailsb}\path|sen.wang@griffith.edu.au|
\newcommand{\keywords}[1]{\par\addvspace\baselineskip
\noindent\keywordname\enspace\ignorespaces#1}

\begin{document}

\mainmatter  % start of an individual contribution


    %%%% Lina: %%%%%
    \title{Hybrid Collaborative Recommendation via Semi-AutoEncoder}
    %\title{Hybrid Collaborative Filtering Model via Semi-AutoEncoder}

\titlerunning{Hybrid Collaborative Recommendation via Semi-AutoEncoder}

\author{Shuai Zhang$^{1}$%
% \thanks{Please note that the LNCS Editorial assumes that all authors have used
% the western naming convention, with given names preceding surnames. This determines
% the structure of the names in the running heads and the author index.}%
\and Lina Yao$^{1}$, Xiwei Xu$^{2}$, Sen Wang$^{3}$ \and Liming Zhu$^{2}$}
\authorrunning{Zhang et al.}

\institute{$^{1}$School of Computer Science and Engineering, University of New South Wales\\
\mailsa\\
$^{2}$Data61, CSIRO\\
\mailsc\\
$^{3}$School of Information and Communication Technology, Griffith University\\
\mailsb\\}


%%% \toctitle{Lecture Notes in Computer Science}
% \tocauthor{Authors' Instructions}
\maketitle

\begin{abstract}
In this paper, we present a novel structure, Semi-AutoEncoder, based on AutoEncoder. We generalize it into a hybrid collaborative filtering model for rating prediction as well as personalized top-$n$ recommendations. Experimental results on two real-world datasets demonstrate its state-of-the-art performances.

\end{abstract}

\section{Introduction}

In the world of exponentially increasing digital data, we need to guide users proactively and provide a new scheme for users to navigate the world. Recommender System (RS) is one of the most effective solutions which help to deliver personalized services or products and overcome information overload. However, traditional recommender systems suffer from the sparseness problem of the rating matrix and are unable to capture the non-linear characteristics of user-item interactions. Here, we propose a hybrid collaborative filtering model based on a novel AutoEncoder structure. It leverages both content information and the learned non-linear characteristics to produce personalized recommendations.
Our contributions are highlighted as follows:

\begin{itemize}
    \item We propose a new AutoEncoder framework named Semi-AutoEncoder. It incorporates side information to assist in learning semantic rich representations or reconstructions flexibly;
    \item We generalize Semi-AutoEncoder into a hybrid collaborative filtering framework to predict ratings as well as generate personalized top-$n$ recommendations;
    \item The experimental results conducted on two public datasets demonstrate that our model outperforms the state-of-the-art methods. We make our implementation publicly available for reproducing the results \footnote{https://github.com/cheungdaven/semi-ae-recsys}.

\end{itemize}

\section{Related Work}
Recent researches have demonstrated the effectiveness of applying AutoEncoder to recommender systems~\cite{deeprecsurvey}. These works can be classified into two categories. The first category aims to use AutoEncoder to learn salient feature representations and integrate them into traditional recommendation models. For example, Li et al.~\cite{li2015deep} designed a model that combines AutoEncoder with probabilistic matrix factorization. Zhang et al.~\cite{autosvd} proposed the AutoSVD++ algorithm which utilizes the features learned by contractive AutoEncoder and the implicit feedback captured by SVD++ to improve the recommendation accuracy. The second category (e.g. AutoRec~\cite{autorec} and \cite{autocl}) focuses on devising recommendation model solely based on AutoEncoder without any help from traditional recommendation models, but these methods do not consider any content information of users and items.

\section{Preliminary}
% Before we dive into the details of our proposed model, we first discuss the preliminaries that we need for building our proposed model.
% In general, most entries of the user-item matrix are missing. The aim of RS is to fill in these blanks.
\subsection{Problem Definition}
Given $N$ items and $M$ users, $R \in \textbf{R}^{N \times M}$ is the rating matrix, and $r^{ui} \in R$ is the rating to item $i \in \{1,...,N\}$ given by user $u \in \{1,...,M\}$. Here, we adopt a partial observed vector $\textbf{r}^u = \{r^{u1},...,r^{uN}\}$, columns of the rating matrix, to represent each user $u$, and partial observed vector $\textbf{r}^i = \{r^{1i},...,r^{Mi}\}$, rows of the rating matrix, to represent each item $i$. For convenience, we use $\textbf{r}^U \in \textbf{R}^{M \times N}$ and $\textbf{r}^I \in \textbf{R}^{N \times M}$ to denote the partial observed vectors for all users and items respectively. In most cases, the ratings can be explicit integer values with the range [1-5] or implicit binary values $\{0, 1\}$, where $0$ means \textit{dislike} and $1$ represents \textit{like}.  We define $\Omega$ as the observed ratings set.

%\textcolor{red}{need more explanation, not smoothly transition.Here, we mainly work on implicit feedback}.
% Here, we denote explicit ratings as $r$ and implicit feedback as $y$. We will utilize both these two kinds of feedback in our proposed model.

% It can be considered as a special case of non-recurrent feed-forward neural network. The training process also follows the stochastic gradient descent scheme and use the gradients computed by back-propagation.

\subsection{AutoEncoder}

AutoEncoder is a neural network for unsupervised learning tasks. It can be applied to dimension reduction, efficient coding or generative modeling~\cite{deeplearningbook}. A typical AutoEncoder consists of three layers. The first layer $x \in \textbf{R}^D$ is the input. The second layer, or the bottleneck layer, usually has less code dimension than the input. We denote the second layer as $h \in \textbf{R}^H (H < D)$ and the output layer $x' \in \textbf{R}^D$ as:
\begin{align}
h = g(Wx + b) \\
x' = f(W_1h + b_1)
\end{align}
where $W \in \textbf{R}^{H \times D}$ and  $W_1 \in \textbf{R}^{D \times H}$ are weight matrices, $b \in \textbf{R}^{H}$ and $b_1 \in \textbf{R}^{D}$ are bias terms. $g$ and $f$ are activation functions such as \textit{Identity} or \textit{Sigmoid}.

AutoEncoder is trained to minimize the reconstruction error between $x$ and $x'$. The loss function is formulated as follows:
\begin{equation}
\mathcal{L}(x,x') = \left \| x - x' \right \|^2 = \left \| x - f(W_1g(Wx + b) + b_1) \right \|^2
\end{equation}

To capture informative features and prevent it from learning identity function, various techniques, such as corrupting the input $x$, adding sparsity penalty terms to the loss function~\cite{deeplearningbook} or stacking several layers together to form a deep neural network, have been proposed. In most cases, we care about the bottleneck layer and use it as compact feature representation. While in this recommendation task, we care more about the output layer.

\section{Methodology}
In this section, we introduce the proposed Semi-AutoEncoder, and detail the Semi-AutoEncoder based hybrid collaborative filtering model.

\subsection{Semi-AutoEncoder}

In general, AutoEncoder requires the dimension of input and output layer to be identical. However, we observe that it does not necessarily have to follow this rule strictly, that is to say, the dimension of these two layers can be different, and it will bring some merits that traditional AutoEncoder do not have in the meantime. Compared to traditional AutoEncoder, our proposed model possess the following advantages:
\begin{itemize}
    \item This model can capture different representations and reconstructions flexibly by sampling different subsets from the inputs.
    \item It is convenient for incorporating additional information in the input layer.
\end{itemize}

By breaking the limitation of the output and input dimensionality, we can devise two variants of AutoEncoder. The output layer can be (\textit{case 1}) longer or (\textit{case 2}) shorter then the input layer. In the former case, the output has a larger size than the input layer, which enables it to generate some new elements from the hidden layer. Although the network can be trained properly, it is difficult to give a reasonable interpretation for these generated entries. While in the later scenario, the output layer is meant for reconstructing certain part of the inputs, and we consider the remaining part to be additional information which facilitates learning better representations or reconstructions. We adopt the term Semi-AutoEncoder to denote the second structure. Figure 1 illustrates the two architectures and the right figure is the structure of Semi-AutoEncoder.

Similarly, a basic Semi-AutoEncoder also has three layers. The input layer $ x \in \textbf{R}^{S}$, hidden layer $ h \in \textbf{R}^{H}$, output layer $x' \in \textbf{R}^{D}$, where $H < D < S$. To train the Semi-AutoEncoder, we need to match the output $x'$ with a designated part of input. We extract a subset with the same length of $x'$ from input $x$, and denote it as $sub(x)$. Then, the network is formulated as follows:
\begin{equation}
h = g(Vx + b)
\end{equation}
\begin{equation}
x' = f(V_1h + b_1)
\end{equation}
\begin{equation}
\mathcal{L}(x,x') = \left \| sub(x) - x' \right \|^2
\end{equation}
where $V \in \textbf{R}^{H \times S}$ and $V_1 \in \textbf{R}^{D \times H}$ are weight matrices, $b \in \textbf{R}^{H}$ and $b_1 \in \textbf{R}^{D}$ are bias terms. When computing the loss function, instead of learning a reconstructions to the whole input $x$, it learns a reconstruction to the subset $sub(x)$. %Similar to AutoEncoder, the proposed Semi-AutoEncoder can be trained with stochastic gradient descent algorithm.

Semi-AutoEncoder can be applied to many areas such as extracting image features by adding captions or descriptions of images, or audio signal reconstruction by integrating environment semantics. In the following text, we will investigate its capability on recommender system by incorporating side information. We are aware of the existing Multimodal Deep learning model~\cite{MultimodelDL} which applies deep AutoEncoder to multi-task learning. The authors proposed the Bimodal Deep AutoEncoder in a denoising fashion. The differences between Bimodal Deep AutoEncoder and Semi-AutoEncoder are: 1) The dimensionalities of the input and output of Bimodal deep AutoEncoder are the same; 2) It requires to be pre-trained with restricted Boltzmann machine (RBM).

\begin{figure}[!tb]
%\vspace{-2mm}
\begin{center}
\begin{minipage}[t]{5.0cm}
\includegraphics[width=5.0cm]{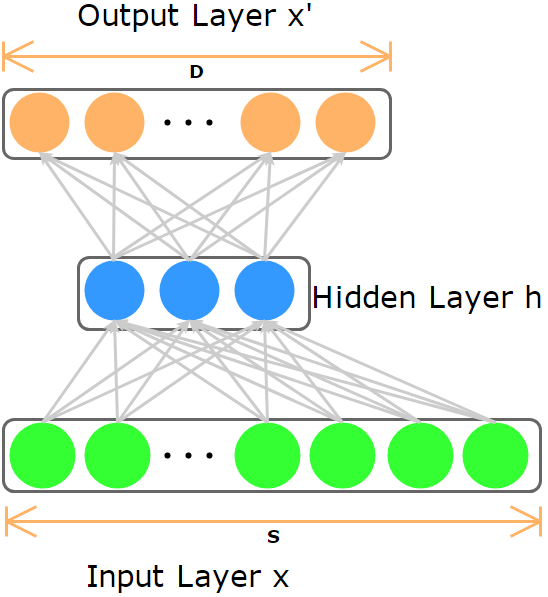}
\centering{(a)}
\end{minipage}
\hspace{12mm}
\begin{minipage}[t]{5.0cm}
\includegraphics[width=5.0cm]{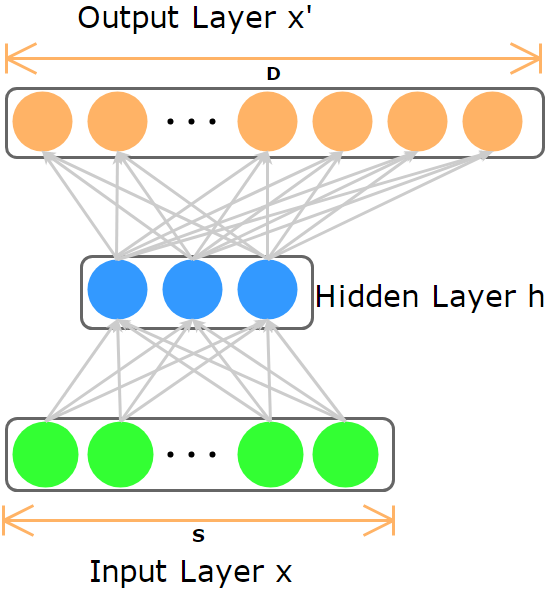}
\centering{(b)}
\end{minipage}
\caption{Illustration of two variants of AutoEncoder.}
%\label{fig:windowsize}
\label{fig:observation}
\end{center}
\vspace{-6mm}
\end{figure}

% The collaborative filtering model based on AutoEncoder focuses more on the output rather than the bottleneck layer. As the figure 1(b) shows, the input is the partial observed vectors $r^i$ or $r^u$. And the output will fill in these blanks of the partial observed vectors. Once either the $r^i$ or $r^u$ being complete, the user-item matrix is complete. Then, we can generate recommendation according the filled values.
% However, this method does not consider any content information of items and users.
% Former Collaborative filtering model based on AutoEncoder such as AutoRec~\cite{autorec} dose not consider any content information of items or users. It cannot provide personalize recommendation and cope with the sparseness of user-item matrix. In this paper,

\subsection{Semi-AutoEncoder for Recommendation}

In this section, we will demonstrate how the proposed Semi-AutoEncoder improve the performance of recommender system regarding two recommendation tasks: rating prediction and ranking prediction. Many existing works are intended to solve one of these problem. Our hybrid model based on Semi-AutoEncoder can solve both of them. Structure of our model is shown in Figure 2. It takes the advantages of Semi-AutoEncoder to incorporate user profiles and item features into collaborative filtering seamlessly.

\subsubsection{Ranking Prediction}

We use $c^u \in \textbf{R}^{K} (u=1,...,M)$ to denote the profile of user $u$. For each user $u$, we have a partial observed vector $\textbf{r}^u$ and a profile vector $c^u$. We concatenate these two vectors together and denote it as $cat(\textbf{r}^u; c^u) \in \textbf{R}^{N+K} (u=1,...,M)$:
\begin{equation}
cat(\textbf{r}^u; c^u) \stackrel{def}{=} \textnormal{concatenation of }  \textbf{r}^u  \ \textnormal{and} \  c^u
\end{equation}

We adopt the uppercase letter $C^U \in \textbf{R}^{M \times K} $ to represent the profiles for all $M$ users, and $cat(\textbf{r}^U;C^U) \in \textbf{R}^{M \times (N+K)}$ to represent the concatenated vectors of all users.
Then, we use the concatenated vector as input and get the hidden representation $h$.
\begin{equation}
h(\textbf{r}^U; C^U) = g(cat(\textbf{r}^U;C^U) \cdot Q  + p)
\end{equation}

% \begin{figure}
% \includegraphics[width=0.6\textwidth]{structuretopn}
% \caption{}
% \end{figure}

\begin{figure}[!tb]
%\vspace{-2mm}
\begin{center}
\begin{minipage}[t]{6cm}
\includegraphics[width=6cm]{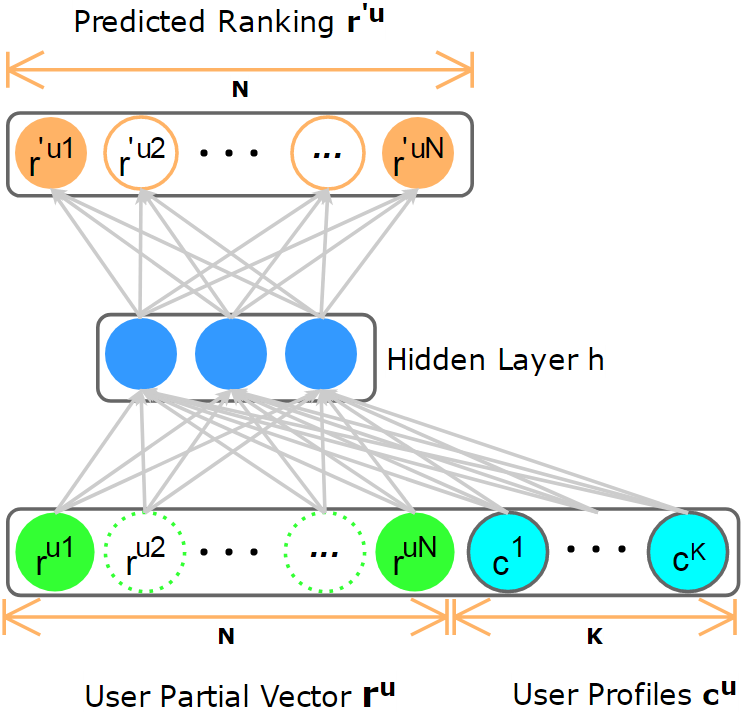}
\centering{(a)}
\end{minipage}
\begin{minipage}[t]{6cm}
\includegraphics[width=6cm]{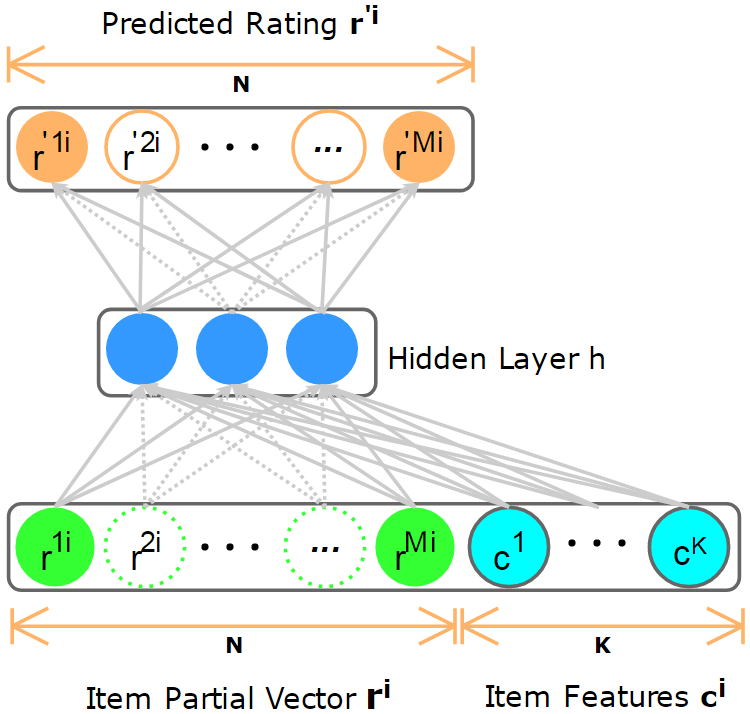}
\centering{(b)}
\end{minipage}
\caption{ Illustration of the hybrid collaborative filtering model based on Semi-AutoEncoder: (a) Ranking prediction; (b) Rating prediction.}
%\label{fig:windowsize}
\label{fig:observation}
\end{center}
\vspace{-6mm}
\end{figure}
Here, $Q \in \textbf{R}^{(N+K) \times H}$ is the weight matrix. $p \in \textbf{R}^H$ is the bias term. Function $g$ is the activation. After getting the hidden representation, we need to reconstruct the input:

\begin{equation}
\textbf{r}^{'U} = f(h(\textbf{r}^U; C^U) \cdot Q_1  + p_1)
\end{equation}

$Q_1 \in \textbf{R}^{H \times N}$ is the reconstruction weight matrix, and $p_1 \in \textbf{R}^N$ is a bias term. $f$ is the activation function. Our goal is to learn an approximate reconstruction to the subset of inputs, where $ sub(x) = \textbf{r}^{u}$, and minimize the differences between $\textbf{r}^{u}$ and reconstruction $\textbf{r}^{'u}$.
% \begin{equation}
% \mathcal{L}(\textbf{r}^{U},\textbf{r}^{'U}) = \left \| \textbf{r}^{U} - \textbf{r}^{'U} \right \|^2_2
% \end{equation}
% = \sum_{u=1}^{u=M} \left \| r^u - r^{'u} \right \|^2_2

To avoid over-fitting, we regularize the weight matrix with $\ell_2$ norm. Finally, the objective function is as follows, and it can be solved by stochastic gradient descent (SGD) algorithm:
\begin{equation}
\begin{aligned}
\underset{Q, Q_1, p, p_1}{\textbf{arg min}} \frac{1}{M}\sum_{u=1}^M \left \| \textbf{r}^{u} - \textbf{r}^{'u} \right \|^2_2  + \frac{  \gamma }{2} (\left \| Q \right \|^2_2 + \left \| Q_1 \right \|^2_2  )
% \\ =  \underset{Q, Q', p, p'}{\textbf{arg min}} \left \| \textbf{r}^{U} - f(Q' \cdot g(Q \cdot cat(\textbf{r}^{U};C^U) + p) + p') \right \|^2  + \frac{  \gamma }{2} (\left \| Q \right \|^2_2 + \left \| Q' \right \|^2_2  )
\end{aligned}
\end{equation}
% \underset{Q, Q', p, p'}{\textbf{arg min}} \sum_{u=1}^{u=M}L(r^u,r^{'u}) + \frac{  \gamma }{2} (\left \| Q \right \|^2_2 + \left \| Q' \right \|^2_2  )
% \\ =  \underset{Q, Q', p, p'}{\textbf{arg min}} \sum_{u=1}^{u=M}\left \| r^u - f(Q' \cdot g(Q \cdot cat(r^u;c^u) + p) + p') \right \|^2_2  + \frac{  \gamma }{2} (\left \| Q \right \|^2_2 + \left \| Q' \right \|^2_2  )

% We also test several other optimization algorithms such as RMSprop and Adam, we present the comparison in section 4.

\subsubsection{Rating Prediction}
The process of predicting ratings is similar to ranking prediction. The differences are: 1) We use item partial observed vector $\textbf{r}^i$ with explicit ratings as inputs; 2) Optimization is only implemented on observed ratings; 3) We integrate item features as additional inputs. Details are as follows.

Similar to ranking prediction, we use $C^I \in \textbf{R}^{N \times K}$ to represent item features of all N items, and $cat(\textbf{r}^I; C^I) \in \textbf{R}^{N \times (M + K)}$ to represent the concatenated vectors, then:
\begin{equation}
h(\textbf{r}^I; C^I) = g( cat(\textbf{r}^I;C^I) \cdot Q + p)
\end{equation}
\begin{equation}
\textbf{r}^{'I} = f(h(\textbf{r}^I; C^I) \cdot Q_1 + p_1)
\end{equation}
% \begin{equation}
% \mathcal{L}(\textbf{r}^{I},\textbf{r}^{'I}) = \left \| \textbf{r}^{I} - \textbf{r}^{'I} \right \|^2_2
% \end{equation}

Where $Q \in \textbf{R}^{(M+K) \times H}, Q_1 \in \textbf{R}^{H \times M}$ are weight matrices, $p \in \textbf{R}^H, p_1 \in \textbf{R}^M$ are bias vectors. The main change for rating prediction is that we only consider observed ratings when updating parameters, thus, the objective function is formulated as below:

\begin{equation}
\underset{Q, Q_1, p, p_1, r^{ui} \in \Omega }{\textbf{arg min}} \frac{1}{N}\sum_{i=1}^N\left \| \textbf{r}^{i} - \textbf{r}^{'i} \right \|^2_2 + \frac{  \gamma }{2} (\left \| Q \right \|^2_2 + \left \| Q_1 \right \|^2_2  )
\end{equation}

We can also deploy the SGD algorithm to learn the parameters. According to our experiments, Adam algorithm is preferred due to its faster convergence.

\section{Experiments}

\subsection{Datasets and Evaluation Metrics}
We conduct experiments on two real-world datasets of different size and density: Movielens 100K and Movielens 1M~\footnote{https://grouplens.org/datasets/movielens/}. For rating prediction, item features consist of genre, year of release. We evaluate the predicted ratings with the widely used metric: Root Mean Square Error (RMSE)~\cite{ricci2011introduction}. We evaluate our model with different training percentages by randomly sampling 50\% and 90\% of rating records as training set, and leaving the remaining part as test set; For ranking prediction, user profiles are made up of age, occupation and gender. Our model aims to predict top $n$ items that the user like most and evaluates against the test data. Same as~\cite{cdl,bprmf,slim}, we use recall to evaluate the performance of ranking quality by randomly choosing 30\% and 50\% of rating records as training set.

Ratings in both datasets are explicit with the range [1-5]. In ranking prediction, we treat ratings that greater than 4 as 1 (\textit{like}) and 0 (\textit{dislike}) for others, and use the explicit ratings for rating prediction.

\subsection{Evaluation Results}

\subsubsection{Rating Prediction} We compare our method with several baselines (e.g., ItemKNN, NMF, PMF, SVD++ etc.) and start-of-the-art deep learning based methods listed below:
\begin{itemize}
\item \textbf{I-RBM} \cite{rbm}, RBM-CF is a generative, probabilistic collaborative filtering model based on restricted Boltzmann machines.
\item \textbf{NN-CF} \cite{hrecauto}, NN-CF is hybrid recommender system built on denoising AutoEncoder.
\item \textbf{mSDA-CF} \cite{li2015deep} , mSDA-CF is a model that combines PMF with marginalized denoising stacked auto-encoders.
\item \textbf{U-AutoRec} \cite{autorec}, U-AutoRec is also a collaborative filtering model based on the AutoEncoder.
\end{itemize}

We decide the hyper-parameters with cross-validation and set learning rate to 0.001, and regularization rate $\gamma=0.1$. We tested different hidden neural size of Semi-AutoEncoder, and set hidden neural size to 500.

% \usepackage{multirow}
\begin{table}[]
\centering
\caption{Average RMSE for Movielens-100k and Movielens-1M with different training data percentages}
\label{my-label}
\begin{tabular}{cccccc}
\toprule
\multirow{2}{*}{Methods} & \multicolumn{2}{c}{Movielens-100K} & \multirow{2}{*}{Methods} & \multicolumn{2}{c}{Movielens-1M}
\\& 80\%  & 50\%    &                          & 80\%            & 50\%           \\
\midrule
ItemKNN                  & 0.926            & 0.940           & ItemKNN                  & 0.882           & 0.892          \\
NMF                      & $0.963\pm0.001$    & $0.994\pm0.005$   & NMF                      & $0.917\pm0.002$    & $0.927\pm0.001$  \\
PMF                      & $0.919\pm0.005$    & $0.951\pm0.002$   & PMF                      & $0.868\pm0.002$   & $0.887\pm0.002$  \\
SVD++                    & $0.946\pm0.001$    & $0.963\pm0.001$  & I-RBM                    & $0.880\pm0.001$   & $0.901\pm0.002$  \\
BMFSI                    & $0.906\pm 0.003$   & $0.933\pm0.003$   & U-AutoRec                & $0.889\pm0.001$   & $0.911\pm0.001$  \\
mSDA-CF                  & $0.902\pm0.003$    & $0.931\pm0.002$   & NN-CF                    & $0.875\pm0.002$   & $0.896\pm0.001$  \\
\textbf{Ours}            & $\textbf{0.896}\pm\textbf{0.003}$    & $\textbf{0.926}\pm\textbf{0.002 }$  & \textbf{Ours}            & $\textbf{0.858}\pm\textbf{0.001}$   & $\textbf{0.882}\pm\textbf{0.001}$ \\
\bottomrule
\end{tabular}
\end{table}

The last three methods are closely relevant to our work. The differences between our proposed model and these models are: 1) Our model is based on the proposed Semi-AutoEncoder, while these model are built on traditional AutoEncoder; 2) Our method is capable of performing both rating and ranking prediction. We incorporate user profiles into the Semi-AutoEncoder tightly to generate personalized top-$n$ recommendation, while these three methods can only predict ratings.

From the experimental results, we can clearly observe that our model beats all the comparison methods in terms of rating prediction.

\subsubsection{Ranking Prediction}
We compare our model with a set of baselines below. To make a fair comparison, we also specify the critical parameters which achieve best performances for each model:

\begin{itemize}
    \item \textbf{MostPopular}, This method recommend the most popular items to users, and the items are weighted by the frequency that they have been seen in the past. It is worth mentioning that it is a deterministic algorithm.
    % \item ItemKNN/UserKNN, Item-based (User-based) K-nearest neighbors method predict recommendation list based on ratings of the k most similar items (users). Here, we set k to 80.
    \item \textbf{BPRMF}~\cite{bprmf}, BPRMF is a matrix-factorization based top-$n$ recommendation model, which mainly focuses on implicit feedback. we set the number of factors to 20, and learning rate to 0.05.

    \item \textbf{SLIM}~\cite{slim}, SLIM is a state-of-the-art top-$n$ recommendation model. We optimize the objective function in a Bayesian personalized ranking criterion. We set the learning rate to 0.05.

\end{itemize}

\begin{table}[]
\centering
\caption{Recall Comparison on Movielens-100K with different training data percentages.}
\label{my-label}
\begin{tabular}{ccccc}
\toprule
\multirow{2}{*}{Methods} & \multicolumn{2}{c}{Training Size 30\%} & \multicolumn{2}{c}{Training Size 50\%} \\
                         & Recall@5 \%        & Recall@10 \%      & Recall@5 \%       & Recall@10 \%       \\
\midrule
MostPopular              & 7.036              & 11.297            & 7.535             & 13.185             \\
BPRMF                    & $9.091\pm0.448$      & $13.736\pm0.246$    & $8.671\pm0.452$     & $13.868\pm0.652$     \\
SLIM                     & $7.051\pm0.286$      & $10.621\pm0.534$    & $8.836\pm0.231$     & $14.334\pm0.118$    \\
\textbf{Ours}            & $\textbf{9.487}\pm\textbf{0.182}$      & $\textbf{14.836}\pm\textbf{0.209}$    & $\textbf{9.543}\pm\textbf{0.365}$     & $\textbf{15.909}\pm\textbf{0.468}$     \\
\textbf{Improvement}       & \textbf{4.355\%}     & \textbf{8.008\%}    & \textbf{8.001\%}    & \textbf{10.987\%}  \\
\bottomrule
\end{tabular}
\end{table}

We set the learning rate to 0.001, regularization rate $\gamma=0.1$, and hidden neural size to 10.  Table 2 highlights the performances of our model and compared methods, and shows that our model outperforms the compared methods by a large margin.

%25 for movielens 1M and

% \begin{table}[]
% \centering
% \caption{Comparision of Recall@10 with four possible activation function combination.}
%     \begin{tabular}{ccc}
% \toprule
% \textbf{f} & \textbf{g} & Recall@10 (\%) \\
% \midrule
% Identity   & Identity   & 15.180\pm0.625 \\
% Sigmoid    & Identity   & 13.442\pm0.732 \\
% Sigmoid    & Sigmoid    & 11.754\pm0.192 \\
% Identity   & Sigmoid    & 15.909\pm0.468 \\
% \bottomrule
% \end{tabular}
% \end{table}
% \noindent
% \begin{minipage}{1\textwidth}
% \begin{minipage}[b]{0.49\textwidth}
%     \centering
%     \includegraphics[width=5cm]{optimizer}
%     \captionof{figure}{Comparison of average test RMSE with different optimization methods on Movielens-100K}
% \end{minipage}
% \hfill
% \begin{minipage}[b]{0.49\textwidth}
%     \centering
%     \begin{tabular}{ccc}
% \toprule
% \textbf{f} & \textbf{g} & Recall@10 (\%) \\
% \midrule
% Identity   & Identity   & 15.180\pm0.625 \\
% Sigmoid    & Identity   & 13.442\pm0.732 \\
% Sigmoid    & Sigmoid    & 11.754\pm0.192 \\
% Identity   & Sigmoid    & 15.909\pm0.468 \\
% \bottomrule
% \\
% \end{tabular}
% \captionof{table}{Comparision of Recall@10 with four possible activation function combination}
% \end{minipage}
% \end{minipage}

% Figure 3 shows the comparison of rating prediction performance with different optimization algorithms.

We compared the performances of different optimization algorithms, Adam, RMSProp and Gradient Descent, and found that Adam converged faster and achieved the rating results. While, for ranking prediction, Gradient Descent performed better than other optimization methods. Besides, activation function $f$ and $g$ can be $Sigmoid, Identity, Relu, Tanh$ etc., In this paper, we mainly investigated $Sigmoid$ and $Identity$. We observed that the combination: $g:Sigmoid; f:Identity$ achieved the best performance both for ranking and rating prediction.

\section{Conclusion and Future Work}
In this paper, we introduce a novel Semi-AutoEncoder structure, and design a hybrid collaborative filtering recommendation model on top of it. We conduct experiments on two real-world datasets and demonstrate that our model outperforms the compared methods. For future work, we plan to extend our proposed model to deep neural network paradigm by integrating more neural layers. We will also consider incorporating richer features such as implicit feedback via Semi-AutoEncoder. In addition, we will conduct experiments to evaluate the impact of Semi-AutoEncoder in other fields such as multi-modal learning or cross-domain recommendation.
%%% too much. %%%
%Moreover, we will improve our model to solve the cold-start problem in recommendation.

\end{document}